\documentclass[11pt]{article}
\usepackage{moriond,epsfig}

\def\be{\begin{equation}}
\def\ee{\end{equation}}
\def\bea{\begin{eqnarray}}
\def\eea{\end{eqnarray}}

\begin{document}
\hfill{CLNS 04/1875}       % for CLNS notes
\vspace*{3.9cm}
%\vspace*{4cm}
\title{The CLEO-c Research Program}

\author{David Asner - for the CLEO collaboration}

\address{University of Pittsburgh, Department of Physics and Astronomy, 3951 O'Hara St, Pittsburgh PA, 15260, USA}

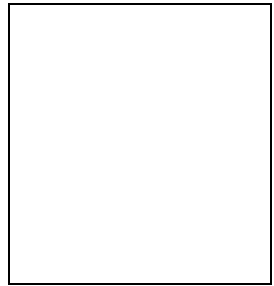
\maketitle\abstracts{
The CLEO-c research program will include studies of leptonic,
semileptonic
and hadronic charm decays, searches for exotic and gluonic matter, and test for
physics beyond the Standard Model. 
The experiment and the CESR 
accelerator were modified to efficiently operate at center-of-mass energies between 3 and 5~GeV.
Data at the $\psi(3770)$ resonance were recorded with 
the CLEO-c detector in September 2003 beginning a new era in the 
exploration 
of the charm sector.
}

\section{Introduction}

The CLEO-c physics program~\cite{bib:cleoc} includes a variety of measurements
that will improve the understanding of Standard Model processes
as well as provide the opportunity to probe physics that lies beyond the
Standard Model. 
The primary components of this program are measurement of absolute
branching ratios for charm mesons with a precision of the order of $1-2\%$,
%(depending upon the mode)
determination of charm meson decay constants and of
the CKM matrix elements $\mid V_{cs} \mid$ 
and $\mid V_{cd} \mid$ at the $1 - 2\%$ level and
investigation of processes in charm decays that are
highly suppressed within the Standard Model. 
A 10 nb cross section for $e^+e^- \to D{\overline D}$ is assumed throughout ref.~\cite{bib:cleoc}.
%Hence, a reconfigured CESR
%electron-positron collider operating at a center of mass energy range between
%3 and 5~GeV together with the CLEO detector will significantly enhance
%our understanding of fundamental Standard Model properties.

%\section{Run Plan and Data Sets}

Since 2003, the CESR accelerator has operated at
center-of-mass energies corresponding to  $\sqrt{s}
\sim 3770$~MeV ($\psi^{\prime\prime}$), $\sqrt{s} \sim 4140$~MeV and $\sqrt{s} \sim 3100$~MeV ($J/\psi$).
The design luminosity at these energies ranges from $5 \times 10^{32}
cm^{-2} s^{-1}$ down to about $1 \times 10^{32} cm^{-2} s^{-1}$ 
yielding 3~fb$^{-1}$ each at the $\psi^{\prime\prime}$ and at
$\sqrt{s} \sim 4140$~MeV above $D_s \bar{D_s}$ threshold and 1~fb$^{-1}$ at the
$J/\psi$. These integrated luminosities correspond to samples of 1.5 million
$D_s \bar{D_s}$ pairs, 30 million $D \bar{D}$ pairs and one billion $J/\psi$
decays~\cite{bib:cleoc}. These datasets will exceed those of the
BESII (Mark III) experiment by factors of 130 (480), 110 (310) and 20 (170), respectively. Additionally, CLEO-c has much better photon energy resolution and particle identification than the BESII and Mark III experiments.
%Table \ref{tab:runplan} summarizes the run plan.
%\newpage

\par
From fall 2001 to spring 2003 CLEO collected a total of 4~fb$^{-1}$ of
data on the $\Upsilon(1S)$, $\Upsilon(2S)$, $\Upsilon(3S)$ and $\Upsilon(5S)$
which is currently under analysis.
These data samples will
increase the available $b\bar{b}$ bound state data by more than an order of
magnitude.

\begin{table}
\caption{The 3-year CLEO-c run plan~$^1$} %~\cite{bib:cleoc}}
\label{tab:runplan}
\begin{center}
\begin{tabular}{lll}
\hline
%    \multicolumn{1}{l}{\bf Year} &
\multicolumn{1}{l}{\bf Resonance}
  & \multicolumn{1}{l}{\bf Anticipated}
  & \multicolumn{1}{l}{\bf Reconstructed}\\
&  \multicolumn{1}{l}{\bf Luminosity} &  \multicolumn{1}{l}{\bf Events}   \\
\hline
 $\psi(3770)$             & $\sim$ 3~fb$^{-1}$ & 30M $D\bar{D}$\\
 $\sqrt{s} \sim 4140$~MeV & $\sim$ 3~fb$^{-1}$ & 1.5M $D_s\bar{D_s}$\\
 $\psi(3100)$             & $\sim$ 1~fb$^{-1}$ & 60M radiative $J/\psi$  \\
\hline
\end{tabular}
\end{center}
\end{table}

%\section{Hardware Requirements}
Only modest hardware modifications are required for low energy operation.
The transverse cooling of the CESR beams will be enhanced by 16 meters of superconducting wiggler magnets. Half of the full complement of 12 wigglers were installed 
in summer 2003 with the additional 6 wigglers scheduled for installation in 2004. 
The CLEO III silicon vertex detector was replaced by a
small, low mass inner drift chamber.
%(Figure \ref{fig:chamber}).
The solenoidal field was
reduced from 1.5~T to 1.0~T. No other modifications are planned.
Prior to the installation of the final 6 wigglers, 
CLEO has accumulated 60.3~pb$^{-1}$ at $\psi(3770)$, 3.1~pb$^{-1}$ at $\psi(2S)$, and
21.0~pb$^{-1}$ of continuum data at $\sqrt{s}=3.67$~GeV with the new detector configuration.

%\begin{figure}[ht]
%%  \includegraphics[height=.2\textheight]{driftchamber}
%  \includegraphics[width=.45\textwidth]{driftchamber}
%  \caption{The new CLEO-c inner drift chamber, 
%           replacing the CLEO-IIIsilicon vertex detector}
%  \label{fig:chamber}
%\end{figure}

\section{Physics Program}

The following sections will outline the CLEO-c physics program.
The first section will focus on the Upsilon spectroscopy, the second section
will describe the charm decay program, the third section
will give an overview about the exotic and gluonic matter studies and the last
section will descibe the oportunities to probe 
physics beyond the Standard Model.

\subsection{Upsilon Spectroscopy \label{sec:upsilon}}

The only established $b\bar{b}$ states below $B\bar{B}$ threshold are the three
vector triplet $\Upsilon$ resonances ($^3S_1$) and the six $\chi_b$ and
$\chi'_b$ (two triplets of $^3P_J$) that are accessible from these parent
vectors via E1 radiative transitions.
% (see Figure \ref{fig:Ystates}). 
CLEO will address a
variety of outstanding physics issues with the data samples
at the $\Upsilon(1S)$, $\Upsilon(2S)$ and $\Upsilon(3S)$, 

%\begin{figure}[ht]
%  \includegraphics[height=.4\textheight]{1600800-006}
%  \caption{Approximate levels of the $b\bar{b}$ states. The name associated with
%the spin-parity assignments are $^1S_0 = \eta_b$, $^3S_1 = \Upsilon$,
%$^1P_1 = h_b$ and $^3P_J = \chi_b$ (triplets with J = 0,1,2).}
%  \label{fig:Ystates}
%\end{figure}

%\begin{itemize}

\noindent {\bf Searches for the $\eta_b$ and $h_b$:}
%The $\eta_b$ is the ground state of $b\bar{b}$. 
Most present theories~\cite{bib:EFI01-10} indicate
the best search method for the $\eta_b$ is the hindered M1 transition from the $\Upsilon(3S)$,
with which CLEO might have a signal of $5 \sigma$ significance in 1~fb$^{-1}$
of data. In the case of the $h_b$, CLEO established an upper limit of
${\cal B}(\Upsilon(3S) \rightarrow \pi^+ \pi^- h_b)$ $<$ 0.18\% at 90\%
confidence level~\cite{bib:hb}. This result, based on $\sim 110$~pb$^{-1}$, 
%already
tests some theoretical predictions~\cite{bib:hb2,bib:hb3,bib:hb4} for this transition which range
from $< 0.01 - 1.0\%$. 
%The resonance run program will measure the mass of the $h_b$ to $\sim 5$~MeV for transition rates $>0.1\%$.

\noindent {\bf Observation of $1^3D_J$ states:}
The $b\bar{b}$ system is unique as it has states with L = 2 that lie below
the open-flavor threshold. These states have been of considerable theoretical
interest, as indicated by many predictions of the center-of-gravity of the
triplet and by a recent review~\cite{bib:EFI01-14}. 
In an analysis of the $\Upsilon(3S)$ CLEO data sample the 
$\Upsilon(1^3D_2)$ state could already be observed in the
four-photon cascade
$\Upsilon(3S) \rightarrow \gamma_1\chi'_b
\rightarrow \gamma_1\gamma_2 \Upsilon(^3D_J) \rightarrow \gamma_1\gamma_2\gamma_3
\chi_b \rightarrow \gamma_1\gamma_2\gamma_3\gamma_4\ell^+\ell^-$.
The mass of the $\Upsilon(1^3D_2)$ state is determined to be
$10161.1 \pm 0.6 \pm 1.6$~MeV/$c^2$~\cite{bib:lp2003}.

\noindent {\bf Observation of New Hadronic $\Upsilon$ Decays:}
Previously, the only hadronic decays of bottomonia experimentally observed were the $\pi\pi$ transitions among vector bottomonium states~\cite{bib:hb}. 
In an analysis of the $\Upsilon(3S)$ CLEO data sample the transitions
$\Upsilon(3S)\to\gamma \chi_b(2P)_{1,2}\to \gamma(\omega\Upsilon(1S))\to\gamma(\pi^+\pi^-\pi^0)(\ell^+\ell^-)$ have been observed. The branching ratios $B(\chi_{b1}\to \omega\Upsilon(1S))$ and  $B(\chi_{b2}\to \omega\Upsilon(1S))$ are 
$(1.63\!^{+0.35}_{-0.31}\!^{+0.16}_{-0.15})\%$ and $(1.10\!^{+0.32}_{-0.28}\!^{+0.11}_{-0.10})\%$, respectively and the ratio of branching ratios is determined to be $0.67^{+0.30}_{-0.22}$~\cite{bib:lp2003}.

\noindent {\bf Glueball candidates in radiative $\Upsilon(1S)$ decays:}
Signals for glueball candidates
in radiative $J/\psi$ decay - a glue-rich environment - might be observed in
radiative $\Upsilon(1S)$ decays.
Naively
one would expect the exclusive radiative decay to be suppressed in $\Upsilon$
decay by a factor of roughly 40, which implies product branching fractions for
$\Upsilon$ radiative decay of $\sim 10^{-6}$. With 1~fb$^{-1}$ of data and
efficiencies of around 30\% one can expect $\sim$ 10 events in each of the
exclusive channels, which would be an important confirmation of the $J/\psi$
studies.

\subsection{Charm Decays \label{sec:charm}}
The observable properties of the charm mesons are determined by the strong
and weak interactions. As a result, charm mesons can be used as a laboratory
for the studies of these two fundamental forces. Threshold charm experiments
permit a series of measurements that enable direct study of the weak
interactions of the charm quark, as well as tests of our theoretical technology
for handling the strong interactions.

%\subsubsection{Leptonic Charm Decays}
\noindent{\bf Leptonic Charm Decays:}
Measurements of leptonic decays in CLEO-c will benefit from the fully 
tagged $D^+$ and $D_s$ decays available at the $\psi(3770)$ and at 
$\sqrt{s} \sim 4140$~MeV. The leptonic decays $D_s \rightarrow \mu\nu$
are detected in tagged events by observing a single charged track of the
correct sign, missing energy, and a complete accounting of the residual
energy in the calorimeter. The clear definition of the initial state, the
cleanliness of the tag reconstruction, and the absence of additional 
fragmentation tracks make this measurement straightforward and essentially
background-free. This will enable measurements of the poorly known
leptonic decay rates for $D^+$ and $D^+_s$ to a precision of 3 - 4\% and will allow
the validation of theoretical calculations of the decay constants
$f_{D}$ and $f_{D_s}$ at the 1 - 2 \% level. 
Table \ref{tab:decayconsts} summarizes
the expected precision in the decay constant measurements.

\begin{table}[ht]
\caption{Expected decay constants errors for leptonic decay modes}
\label{tab:decayconsts}
\begin{center}
\begin{tabular}{lcc}
\hline
    \multicolumn{1}{l}{\bf }
  & \multicolumn{2}{c}{\bf Decay Constant Error \%} \\
    \multicolumn{1}{l}{\bf Decay Mode}
  & \multicolumn{1}{l}{\bf PDG 2000}
  & \multicolumn{1}{l}{\bf CLEO-c~\cite{bib:cleoc}}   \\
\hline
$D^+ \rightarrow \mu^+ \nu$ ($f_D$)     & Upper Limit  & 2.3  \\
$D^+_s \rightarrow \mu^+ \nu$   ($f_{D_s}$) & 17           & 1.7  \\
$D^+_s \rightarrow \tau^+ \nu$  ($f_{D_s}$) & 33           & 1.6  \\
\hline
\end{tabular}
\end{center}
\end{table}

%\subsubsection{Semileptonic Charm Decays}
\noindent{\bf Semileptonic Charm Decays:}
The CLEO-c program will provide a large set of precision measurements in the
charm sector against which the theoretical tools needed to extract CKM matrix
information precisely from heavy quark decay measurements will be tested and
calibrated.
\par
CLEO-c will measure the branching ratios of many exclusive semileptonic modes,
including
$D^0 \rightarrow K^- e^+ \nu$,
$D^0 \rightarrow \pi^- e^+ \nu$,
$D^0 \rightarrow K^{-} e^+ \nu$,
$D^+ \rightarrow \bar{K}^{0} e^+ \nu$,
$D^+ \rightarrow \pi^0 e^+ \nu$,
$D^+ \rightarrow \bar{K}^{0*} e^+ \nu$,
$D^+_s \rightarrow \phi e^+ \nu$ and
$D^+_s \rightarrow \bar{K}^{0*} e^+ \nu$.
The measurement in each case is based on the use of tagged events where the
cleanliness of the environment provides nearly background-free signal samples,
and will lead to the determination of the CKM matrix elements
$\mid V_{cs} \mid$ and $\mid V_{cd} \mid$ with a precision level of
1.6\% and 1.7\%, respectively. Measurements of the vector and axial vector form
factors $V(q^2)$, $A_1(q^2)$ and $A_2(q^2)$ will also be possible at the
$\sim$ 5\% level. Table \ref{tab:semileptonic} summarizes the expected fractional error
on the branching ratios.

\begin{table}[ht]
\caption{Expected branching fractional errors for selected semileptonic decay modes}
\label{tab:semileptonic}
\begin{center}
\begin{tabular}{lcc}
\hline
    \multicolumn{1}{l}{\bf }
  & \multicolumn{2}{c}{\bf BR fractional error \%} \\
    \multicolumn{1}{l}{\bf Decay Mode}
  & \multicolumn{1}{l}{\bf PDG 2000}
  & \multicolumn{1}{l}{\bf CLEO-c~\cite{bib:cleoc}}   \\
\hline
$D^0 \rightarrow K \ell \nu$    & 5        & 0.4                       \\
$D^0 \rightarrow \pi \ell \nu$  & 16       & 1.0                      \\
$D^+ \rightarrow \pi \ell \nu$  & 48       & 2.0                      \\
$D_s \rightarrow \phi \ell \nu$ & 25       & 3.1                       \\
\hline
\end{tabular}
\end{center}
\end{table}

HQET provides a successful description of the lifetimes of charm hadrons and
of the absolute semileptonic branching ratios of the $D^0$ and $D_s$
~\cite{bib:bcp3}. Isospin invariance in the strong forces implies
$\Gamma_{SL}(D^0) \simeq \Gamma_{SL}(D^+)$ up to corrections of 
${\cal O}(\tan^2 \theta_C) \simeq 0.05$. Likewise, $SU(3)_{Fl}$ symmetry
relates $\Gamma_{SL}(D^0)$ and $\Gamma_{SL}(D_s^+)$, but a priori would
allow them to differ by as much as 30\%. However, HQET suggests that they
should agree to within a few percent. The charm threshold region 
is the best place to
measure absolute inclusive semileptonic charm branching ratios, in particular
${\cal B}(D_s \rightarrow X \ell\nu)$ and thus $\Gamma_{SL}(D_s)$.

%\subsubsection{Implications of the Leptonic and Semileptonic Measurements for 
%               CKM}
\noindent{\bf Implications for CKM Triangle:}
%Every weak decay involving leptons depends on both CKM elements and on
%hadronic matrix elements. As described in the sections above, CLEO-c data can 
%be used for ca\-li\-brating the theoretical tools that will determine the
%hadronic terms and for extracting the essential CKM elements. 
%\par
%Combining the leptonic and semileptonic measurements leads to ``direct''
%determinations of the CKM elements $\mid V_{cd} \mid$ and 
%$\mid V_{cs} \mid$. The results are
%shown in Table \ref{tab:VcdVcs}. For this table LatticeQCD is assumed
%and validated across a wide range of charm and onium decay measurements,
%to which CLEO-c will provide decay constants with 1\% accuracy.
%
%\begin{table}[ht]
%\begin{tabular}{lcc}
%\hline
%    \multicolumn{1}{l}{\bf Decay Mode}
%  & \multicolumn{1}{c}{\bf CKM Element}
%  & \multicolumn{1}{c}{\bf CKM Precision}   \\
%\hline
%$D_s \rightarrow \mu^+ \nu$     & $\mid V_{cs} \mid$ & 1.7  \\
%$D_s \rightarrow \tau^+ \nu$    & $\mid V_{cs} \mid$ & 1.6  \\
%$D^0 \rightarrow K^- e^+ \nu$   & $\mid V_{cs} \mid$ & 1.6  \\
%\hline
%$D^+ \rightarrow \mu^+ \nu$     & $\mid V_{cd} \mid$ & 2.3  \\
%$D^0 \rightarrow \pi^- e^+ \nu$ & $\mid V_{cd} \mid$ & 1.7  \\
%\hline
%\end{tabular}
%\caption{Collected results for $\mid V_{cd} \mid$ and $\mid V_{cs} \mid$}
%\label{tab:VcdVcs}
%\end{table}
%The impact of the entire suite of CLEO-c measurements on the current
%knowledge of the CKM matrix is summarized in the following paragraphs.
The CLEO-c program of leptonic and semileptonic
measurements has two components: one of calibrating and validating
theoretical methods for calculating hadronic matrix elements, which can then be
applied to all problems in CKM extraction in heavy quark physics; and one of
extracting CKM elements directly from the CLEO-c data. The direct results
of CLEO-c are the precise determination of $\mid V_{cd} \mid$, 
$\mid V_{cs} \mid$, $f_D$,
$f_{D_s}$, and the semileptonic form factors. The precision knowledge
of the decay constants $f_D$ and $f_{D_s}$, together with the rigorous
calibration of theoretical techniques for calculating heavy-to-light
semileptonic form factors, are required for the direct extraction of CKM 
elements from CLEO-c. This also drives
the indirect results, namely the precision 
extraction
of CKM elements from experimental measurements of the $B_d$ mixing 
frequency, the $B_s$ mixing frequency, and the $B \rightarrow \pi\ell\nu$
decay rate measurements which will be performed by
BaBar, Belle, CDF, D0, BTeV, LHCb, ATLAS and CMS.
%\par
In Table \ref{tab:newCKM} the combined projections are presented~\cite{bib:cleoc}. In the
determination of the CKM elements $\mid V_{cd} \mid$ 
and $\mid V_{cs} \mid$ from $B$ and $B_s$
mixing $\mid V_{tb} \mid = 1$ is used. The tabulation also includes improvement
in the direct measurement of $\mid V_{tb} \mid$ 
expected from the Tevatron experiments~\cite{bib:tevatron1}. %,bib:tevatron2}.

\begin{table}[ht]
\caption{CKM elements at present and after CLEO-c~$^1$} %~\cite{bib:cleoc}}
\label{tab:newCKM}
\begin{center}
\begin{tabular}{lll}
\hline
    \multicolumn{3}{c}{\bf Present Knowledge $\to$ After CLEO-c} \\
\hline
$\delta V_{ud}/V_{ud}=$ 0.1\% $\to$ 0.1\% & $\delta V_{us}/V_{us}=$ 1\%  $\to$ 1\%  & 
$\delta V_{ub}/V_{ub}=$ 25\%  $\to$ 5\%\\ 
$\delta V_{cd}/V_{cd}=$ 7\%  $\to$ 1\% & $\delta V_{cs}/V_{cs}=$ 16\% $\to$ 1\%  & 
$\delta V_{cb}/V_{cb}=$ 5\%  $\to$ 3\% \\
$\delta V_{td}/V_{td} = 36\%$  $\to$ 5\% & $\delta V_{ts}/V_{ts} = 39\%$  $\to$ 5\% & 
$\delta V_{tb}/V_{tb} = 29\%$  $\to$ 15\% \\ 
\hline
%   \multicolumn{3}{c}{\bf After CLEO-c}   \\
%\hline
%$\delta V_{ud}/V_{ud}=$ 0.1\% &
%$\delta V_{us}/V_{us}=$ 1\%   & $\delta V_{ub}/V_{ub}=$ 5\%   \\
%%$\delta V_{cd}/V_{cd}=$ 1\%   &
%$\delta V_{cs}/V_{cs}=$ 1\%   & $\delta V_{cb}/V_{cb}=$ 3\%   \\
%$\delta V_{td}/V_{td}=$ 5\%   &
%$\delta V_{ts}/V_{ts}=$ 5\%   & $\delta V_{tb}/V_{tb}=$ 15\%  \\
%\hline
\end{tabular}
\end{center}
\end{table}

%\subsubsection{Hadronic Charm Decays}
\noindent{\bf Hadronic Charm Decays:}
%The $D^0$ is the best known of all the charm hadrons. 
The CLEO and ALEPH
experiments by far provide the most precise measurements for the decay
$D^0 \rightarrow K^- \pi^+$. They use the same technique by looking at
$D^{*+} \rightarrow \pi^+ D^0$ decays and taking the ratio of the $D^0$ decays
into $K^- \pi^+$ to the number of decays with only the $\pi^+$ from the $D^{*+}$
decay detected. The dominant systematic uncertainty is the background level in
the latter sample. In both experiments, the systematic errors exceed the
statistical errors. 
\begin{table}[ht]
\caption{Expected branching fractional errors for hadronic decay modes~$^1$} %~\cite{bib:cleoc}}
\label{tab:hadronic}
\begin{center}
\begin{tabular}{lcc}
\hline
    \multicolumn{1}{l}{\bf }
  & \multicolumn{2}{c}{\bf BR fractional error \%} \\
    \multicolumn{1}{l}{\bf Decay Mode}
  & \multicolumn{1}{l}{\bf PDG 2000}
  & \multicolumn{1}{l}{\bf CLEO-c~\cite{bib:cleoc}}   \\
\hline
$D^0 \rightarrow K \pi$    & 2.4      & 0.6                       \\
$D^+ \rightarrow K \pi \pi$  & 7.2      & 0.7                       \\
$D_s \rightarrow \phi \pi$ & 25       & 1.9                       \\
\hline
\end{tabular}
\end{center}
\end{table}
The $D^+$ absolute branching ratios are determined by using fully reconstructed
$D^{*+}$ decays, comparing $\pi^0 D^+$ with $\pi^+ D^0$ and using isospin
symmetry. Hence, this rate cannot be determined any better than the
absolute $D^0$ decay rate using this technique. 
The $D^+_s$ absolute branching ratios are determined by comparing fully reconstructed $B\to D^{(*)}D_s^{*+}$ to the partially reconstructed $B\to D^{(*)}D_s^{*+}$ requiring only the $\gamma$ from the $D_s^{*+}$ decay. Here the dominant systematic uncertainty is due to the background shape in the partially reconstructed sample.
By reconstructing both $D$ mesons in $D{\overline D}$ decays,
%$D^0\bar{D}^0$, $D^+D^-$ and $D_s^+D_s^-$ decays, 
the background can be reduced to almost zero and the branching ratio fractional
error can be improved significantly (see Table \ref{tab:hadronic}).

\subsection{Exotic and Gluonic Matter \label{sec:glue}}
The approximately one billion $J/\psi$ produced at CLEO-c will be a
glue factory to search for glueballs and other glue-rich states via
$J/\psi \rightarrow gg \rightarrow \gamma X$ decays. The region of
$1 < M_X < 3$~GeV/$c^2$ will be explored with partial wave analyses for
evidence of scalar or tensor glueballs, glueball-$q\bar{q}$ mixtures, exotic
quantum numbers, quark-glue hybrids and other new forms of matter predicted by
QCD. This includes the establishment of masses, widths, spin-parity quantum
numbers, decay modes and production mechanisms for any identified states, a
detailed exploration of reported glueball candidates such as
%such as the tensor candidate $f_J(2220)$ 
the scalar states $f_0(1370)$, $f_0(1500)$ and $f_0(1710)$, and
the examination of the inclusive photon spectrum $J/\psi \rightarrow \gamma$X
with $<$ 20 MeV photon resolution and identification of states with up to 100
MeV width and inclusive branching ratios above $1 \times 10^{-4}$. 
%A Monte Carlo study of inclusive radiative $J/\psi$ decays in CLEO-c is
%shown in Figure \ref{fig:inclphoton} based on a sample of 60 million
%$J/\psi$ decays and assuming 
%${\cal B}(J/\psi \rightarrow \gamma f_J(2220)) = 8 \times 10^{-4}$.
%A monochromatic photon line from the $J/\psi \rightarrow \gamma f_J(2220)$
%decay is clearly seen. The signal efficiency is 24\%. With 
%$10^9$ $J/\psi$ decays, CLEO-c will be able to discover any narrow
%resonance produced in radiative $J/\psi$ decays with inclusive branching
%fractions of order $10^{-4}$ or greater.

%\begin{figure}[ht]
%  \includegraphics[height=.4\textheight]{2770301-010}
%  \caption{The inclusive photon spectrum from $J/\psi$ decays from a 
%Monte Carlo simulation in CLEO-c. Signals from $\eta'$, $\eta(1440)$ and
%$f_J(2220)$ are clearly visible. 
%A broad signal from $f_4(2050)$ production is also evident.}
%  \label{fig:inclphoton}
%\end{figure}

In addition,
spectroscopic 
searches for new states of the $b\bar{b}$ system and for exotic
hybrid states such as $cg\bar{c}$ will be made using the 4~fb$^{-1}$
$\Upsilon(1S)$, $\Upsilon(2S)$, $\Upsilon(3S)$ and $\Upsilon(5S)$
data. Analysis of
$\Upsilon(1S) \rightarrow \gamma X$ will play an important role in verifying
any glueball candidates found in the $J/\psi$ data.

\subsection{Charm Beyond the Standard Model}
CLEO-c has the opportunity to probe for physics beyond the Standard
Model. Three highlights - rare charm decays, $D^0-\bar{D}^0$-mixing and $CP$ 
violation 
- are discussed in the following sections.

%\subsubsection{Rare Charm Decays}
\noindent{\bf Rare Charm Decays:}
Rare decays of charmed mesons and baryons provide ``background-free''
probes of new physics effects. In the framework of the Standard Model (SM)
these processes occur only at one loop level. SM predicts vanishingly small
branching ratios for processes such as $D \to \pi/K^{(*)} \ell^+\ell^-$
%because of the absence in the SM of the
%super-heavy bottom-type quark supplemented by 
due to the almost perfect GIM cancellation
between the contributions of strange and down quarks. 
%This is very different
%from the familiar case of bottom quark decays where the top quark contribution
%dominates the decay amplitude. 
This causes the SM predictions for these
transitions to be very uncertain.
%, as the pertubative GIM cancellation mechanism
%is not effective for soft, long-distance contributions. 
In addition, in many
cases annihilation topologies also give sizable contribution. 
%At the end, any
%anomalous enhancement of a given branching ratio would have to be compared
%to the (dominant long-distance) SM amplitude. Fortunately, 
Several
model-dependent estimates exist indicating that the SM predictions for these
processes are still far below current experimental sensitivities~\cite{bib:rare1,bib:rare2}. 
%From there is also
%follows that experiments which can measure rare $D$ decay branching ratios at 
%the level of $10^{-6}$, such as CLEO-c, will start to confront models of new 
%physics in an interesting way.

%\subsubsection{$D\bar{D}$-Mixing}
\noindent{\bf ${\bf D^0-\bar{D}^0}$ Mixing:}
Neutral flavor oscillation in the $D$ meson system is highly suppressed
within the Standard Model. The time evolution of a particle produced as a $D^0$
or ${\overline D}^0$, in the limit of $CP$ conservation, is governed by four parameters:
$x=\Delta m/\Gamma$, $y=\Delta \Gamma/2\Gamma$ characterize the mixing matrix, 
$\delta$ the relative strong phase
between Cabibbo favored (CF) and doubly-Cabibbo suppressed (DCS) amplitudes and 
$R_D$ the DCS decay rate relative to the CF decay rate~\cite{bib:asner}. 
Standard Model based predictions for $x$ and $y$, as well as a variety of non-Standard 
Model expectations, span several orders of magnitude~\cite{bib:Nelson}.
It is reasonable to assume that $x\approx y \approx 10^{-3}$ in the Standard Model.
The mass and width differences $x$ and $y$ can be measured in a variety of ways.
The most precise limits are obtained by exploiting the time-dependence of 
$D$ decays~\cite{bib:asner}. Time-dependent analyses are not feasible at 
CLEO-c; however,
the quantum-coherent $D^0{\overline D}^0$ state provides time-integrated sensitivity 
to $x$, $y$ at ${\cal O}(1\%)$ level and $\cos\delta\sim 0.05$~\cite{bib:cleoc,bib:ddmixing2}. Although CLEO-c does not have sufficient sensitivity to observe Standard
Model charm mixing the projected results compare favorably with current experimental results; 
see Fig.~1 in Ref.~\cite{bib:asner}.

\noindent{\bf ${\bf CP}$ Violation:}
%Standard Model $CP$ violation is strongly suppressed in charm. 
Theoretical predictions for the rate of $CP$ violation in the Standard Model have significant uncertainties.
Standard Model predictions for the rate of $CP$ violation in charm mesons are as large
as 0.1\% for $D^0$ decays and as large as 1\% for certain $D^+$ and $D^+_s$ decays~\cite{bib:Buccella}.
%In addition to indirect $CP$ violation, both SM and new physics effects can 
%induce different contributions to the decay amplitudes of $D$ mesons.
%This phenomenon can be traced back to the appearance of complex-valued
%couplings (CKM parameters) in the $\Delta C = 1$ Lagrangian that mediates $D$
%decays and leads to a $CP$-violating difference between decay rates of 
%$CP$-conjugated states.
%\par

The process
%\begin{displaymath}
$e^+ e^- \rightarrow \psi(3770) \rightarrow D^0 \bar{D^0}$
%\end{displaymath}
produces an eigenstate of $CP+$, in the first step, since the $\psi$(3770) has
$J^{PC}$ equal to $1^{--}$. Now consider the case where both the $D^0$ and
the $\bar{D^0}$ decay into $CP$ eigenstates. Then the decays
%\begin{displaymath}
$\psi(3770) \rightarrow f^i_+ f^j_+ ~~or~~ f^i_- f^j_-$
%\end{displaymath}
are forbidden, where $f_+$ denotes a $CP+$ eigenstate and $f_-$ denotes a $CP-$
eigenstate. This is because
%\begin{displaymath}
$CP(f^i_{\pm} ~f^j_{\pm}) = (-1)^\ell = -1$
%\end{displaymath}
for the $\ell = 1 ~\psi$(3770).
%\par
Hence, if a final state such as ($K^+K^-$)($\pi^+\pi^-$) is observed, one
immediately has evidence of $CP$ violation. Moreover, all $CP+$ and $CP-$ 
eigenstates
can be summed over for this measurement. The expected sensitivity to direct $CP$
violation is $\sim 1\%$.
This measurement can also be performed at higher energies where the final
state $D^{*0} \bar{D^{*0}}$ is produced. When either $D^*$ decays into a
$\pi^0$ and a $D^0$, the situation is the same as above. When the decay is 
$D^{*0} \rightarrow \gamma D^0$ the $CP$ parity is changed by a multiplicative
factor of -1 and all decays $f^i_+ f^j_-$ violate $CP$~\cite{bib:CP}. Additionally, $CP$
asymmetries in $CP$ even initial states depend linearly on $x$ allowing sensitivity to
$CP$ violation in mixing of $\sim 3\%$~\cite{bib:cleoc}.

{\noindent \bf Dalitz Plot Analyses:} A Dalitz plot analysis of multibody final states measures amplitudes and phases
rather than the rates and so may provide greater sensitivity to $CP$ violation.
In the limit of $CP$ conservation, charge conjugate decays will have the same
Dalitz distribution. Although the $D^+$ and $D_s^+$ decays are self-tagging,
there have been no reported Dalitz analyses that search for $CP$ violation with
charged $D$'s. The decay $D^0 \to K_S \pi^+\pi^-$ proceed through intermediate
states that are $CP+$ eigenstates, such as $K_S f_0$, $CP-$ such as $K_S\rho$ and flavor eigenstates such as $K^{*-}\pi^+$~\cite{bib:asner2}. 
It is noteworthy that for uncorrelated $D^0$
the interference between $CP+$ and $CP-$ eigenstates integrates to zero across the 
Dalitz plot but for correlated $D$ the interference between $CP+$ and $CP-$ eigenstates
is locally zero. The Dalitz plots for $\psi(3770) \to D^0{\overline D}^0 \to f_+K_S\pi^+\pi^-$ and  $\psi(3770) \to D^0{\overline D}^0 \to f_-K_S\pi^+\pi^-$ will be
distinct and the Dalitz plot for the untagged sample  $\psi(3770) \to D^0{\overline D}^0 \to X K_S\pi^+\pi^-$ will be distinct from that observed with uncorrelated $D$'s from continuum production at $\sim 10$~GeV~\cite{bib:asner2}.
The sensitivity at CLEO-c to $CP$ violation with Dalitz plot analyses has not yet been evaluated.

\section{Summary}
The high-precision charm and quarkonium data will permit a broad suite of
studies of weak and strong interaction physics as well as probes of
new physics. In the threshold charm sector
measurements are uniquely clean and make possible the unambigous determinations
of physical quantities discussed above. 
The advances in strong interaction calculations enabled by \hbox{CLEO-c} will allow
advances in weak interaction physics in all heavy quark endeavors and in future
explorations for physics beyond the Standard Model.
%CLEO-c will utilize a variety of tools,
%namely $J/\psi$ radiative decays, two-photon collisions (using almost real, as
%well as highly virtual space-like photons), deep inelastic Coulomb scattering
%and continuum production via $e^+ e^-$ annihilation to obtain significant new
%information on the spectrum of hadrons, both normal and exotic, and their decay
%channels. A quantitative improvement can be expected not only from the large
%accumulated statistics, but also from combining the results obtained using all
%these tools together with the results from the $\Upsilon$ resonance runs. The
%significance of this is better sensitivity, reduced systematics and a better
%chance to obtain a coherent picture of the hadron sector.

%\section{Acknowledgments}
%I am delighted to acknowledge the invaluable contributions of many individuals
%to the development of the CLEO-c and CESR-c program and the outstanding
%contributions of my CLEO colleagues over the life of the experiment. The
%experimental aspects of this program are based on their effort and experience.


\section*{References}
\begin{thebibliography}{99}
\bibitem{bib:cleoc}
Briere, R.~A., et~al., \emph{{\rm CLNS-01-1742}} (2001).

\bibitem{bib:EFI01-10}
Godfrey, S., and Rosner, J.~L., \emph{Phys. Rev.}, \textbf{D64}, 074011
  (2001).

\bibitem{bib:hb}
Butler, F., et~al., \emph{Phys. Rev.}, \textbf{D49}, 40--57 (1994).

\bibitem{bib:hb2}
Kuang, Y.-P., and Yan, T.-M., \emph{Phys. Rev.}, \textbf{D24}, 2874 (1981).

\bibitem{bib:hb3}
Voloshin, M.~B., \emph{Sov. J. Nucl. Phys.}, \textbf{43}, 1011 (1986).

\bibitem{bib:hb4}
Voloshin, M.~B., and Zakharov, V.~I., \emph{Phys. Rev. Lett.}, \textbf{45}, 688
  (1980).

\bibitem{bib:EFI01-14}
Godfrey, S., and Rosner, J.~L., \emph{Phys. Rev.}, \textbf{D64}, 097501
  (2001).

\bibitem{bib:lp2003}
Skwarnicki, T., \emph{Proceedings of Lepton-Photon} (2003).

\bibitem{bib:bcp3}
Bigi, I. I.~Y., \emph{Proceedings of BCP3} (2000).

\bibitem{bib:tevatron1}
Swain, J., and Taylor, L., \emph{Phys. Rev.}, \textbf{D58}, 093006 (1998).

\bibitem{bib:rare1}
Fajfer, S., et~al., \emph{Phys. Lett.}, \textbf{B487}, 81--86 (2000).

\bibitem{bib:rare2}
Burdman, G., et~al., \emph{Phys. Rev.}, \textbf{D52}, 6383--6399 (1995).

\bibitem{bib:asner}
Asner, D., \emph{{\rm $D^0-\bar D^0$ Mixing in Review of Particle Physics}}, \emph{Phys. Lett.}, \textbf{B592}, 1 (2004).

\bibitem{bib:Nelson}
%Nelson, H.~N., \emph{{\rm hep-ex/9908021}} (1999).
Petrov, A.~A., \emph{Proceedings of Flavor Physics and $CP$ Violation} (2003).
\bibitem{bib:ddmixing2}
Gronau, M., Grossman, Y., and Rosner, J.~L., \emph{Phys. Lett.}, \textbf{B508},
  37--43 (2001).

\bibitem{bib:Buccella}
Buccella, F., Lusignoli, M., and Pugliese, A., \emph{Phys. Lett.},
  \textbf{B379}, 249--256 (1996).

\bibitem{bib:CP}
Bigi, I. I.~Y., and Sanda, A.~I., \emph{Cambridge Monogr. Part. Phys. Nucl.
  Phys. Cosmol.}, \textbf{9}, p. 180 (2000).

\bibitem{bib:asner2}
Muramatsu, H., et~al., \emph{Phys. Rev. Lett.}, \textbf{89}, 251802 (2002).
\end{thebibliography}
\end{document}